\documentclass[showpacs,aps,pra,twocolumn,superscriptaddress]{revtex4} 
\usepackage{array}
\usepackage{times,amsmath,graphicx,amssymb}
\usepackage{color}
\usepackage{ulem}
\usepackage{soul}

\usepackage{tikz}
\usetikzlibrary{matrix}

\definecolor{grey}{rgb}{.6,.6,.6}
\definecolor{forestgreen}{rgb}{.13,.55,.13}
\definecolor{amber}{rgb}{1.00,.31,.0}

\def \hH{ \hat{\mathcal{H}}}
\def \bse{\begin{subequations}}
\def \ese{\end{subequations}}

\newcommand{\ev}[1]{\ensuremath{\left\langle #1 \right\rangle}}

\newcommand{\IN}{\textup{in}}
\newcommand{\OUT}{\textup{out}}
\newcommand{\COH}{\textup{coh}}

\begin{document}
 
\title{Non-reciprocal quantum interactions and devices via autonomous feed-forward}

\author{A. Metelmann}
\affiliation{Department of Electrical Engineering, Princeton University, Princeton, New Jersey 08544, USA}
\author{A. A. Clerk}
\affiliation{Department of Physics, McGill University, 3600 rue University, Montr\'{e}al, Quebec, H3A 2T8 Canada}

\date{\today}

\begin{abstract}
In a recent work [A. Metelmann and A. A. Clerk, Phys.~Rev.~X {\bf 5}, 021025 (2015)], a general reservoir-engineering approach for generating non-reciprocal quantum interactions and devices was described.  We show here
how in many cases this general recipe can be viewed as an example of autonomous feed-forward: the full dissipative evolution is identical to 
the unconditional evolution in a setup where an observer performs an ideal quantum measurement of one system, and then uses the results to drive a second system.  
We also extend the application of this approach to non-reciprocal quantum amplifiers, showing the added functionality possible when using two engineered reservoirs.  In particular, we demonstrate how to construct an ideal phase-preserving cavity-based amplifier which is full non-reciprocal, quantum-limited and free of any fundamental gain-bandwidth constraint.
\end{abstract}
	
\pacs{42.65.Yj, 03.65.Ta, 42.50.Wk, 07.10.Cm}

\maketitle


\section{Introduction}

Directional photonic devices such as isolators and circulators play a crucial role in numerous settings, ranging from the protection of lasers against spurious reflections, 
to the isolation of superconducting qubits from noise at microwave frequencies.  More fundamentally, the possibility of having one-way photonic interactions and transport could enable a variety of new kinds of quantum photonic states and phases (see, e.g.,~\cite{lodahl_chiral_2016}).  
Standard approaches for achieving non-reciprocity make use of magneto-optical effects, which have the disadvantage of being bulky and requiring large static magnetic fields; this makes on-chip integration difficult, especially in superconducting circuits.  There has thus been considerable interest in finding alternate routes to non-reciprocity that allow greater flexibility.  Many such works ultimately rely on using the relative phases of various control drives applied to a system to effectively break time-reversal symmetry 
\cite{yu_complete_2009, lira_electrically_2012,kamal_noiseless_2011, ranzani_geometric_2014,kamal_asymmetric_2014,ranzani_graph-based_2015}.
As such, there is a strong connection to work examining means for creating synthetic gauge fields for neutral particles like photons \cite{haldane_possible_2008, raghu_analogs_2008, hafezi_robust_2011, fang_realizing_2012} .

\begin{figure} 
  \centering\includegraphics[width=0.48\textwidth]{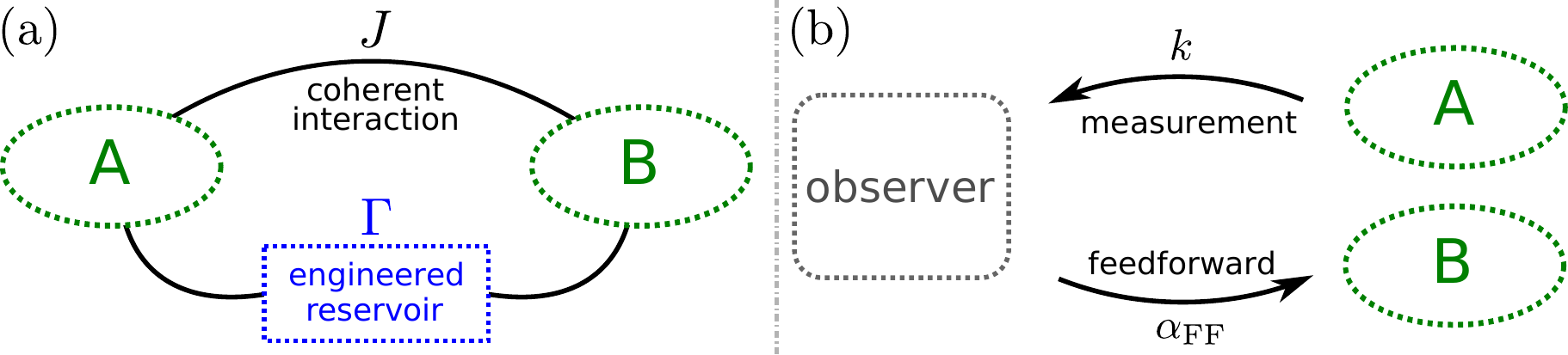} 
 	\caption{(a) Recipe for directionality: System A and B are coupled via a coherent interaction ($J$) and via a dissipative interaction ($\Gamma$) mediated by an engineered reservoir.
 	         By balancing these interactions the system can be rendered uni-directional, e.g. for $J \equiv i\Gamma/2$ system B is driven by system A but not vice versa.
 	        (b) Feed-forward approach to directionality: an observer continuously measures system A and 
            uses the measurement record to apply an appropriate feed-forward force on system B. 
            The resulting dynamics realizes an effective non-reciprocal interaction between
            the systems.  $k$ denotes the rate at which the observer measures system A,  
            while $\alpha_{\rm FF}$ describes the strength of the feed-forward
	        process.  }
 	\label{Fig.:SketchFFScheme}
\end{figure}

Spurred by these motivations, we recently described in Ref.~\onlinecite{metelmann_nonreciprocal_2015} an extremely general method for constructing non-reciprocal interactions and devices using the ideas of reservoir engineering.  By balancing a given Hamiltonian interaction between two systems A and B with its dissipative counterpart (i.e., an interaction mediated by a dissipative reservoir), we demonstrated that almost any starting interaction could be rendered directional (see Fig.~\ref{Fig.:SketchFFScheme}(a)).  This ``recipe" provided a unifying framework for understanding a variety of existing proposals for achieving non-reciprocity.  It also allowed us to formulate a number of new kinds of devices, in particular quantum-limited directional amplifiers based on coupled cavity modes.  The reservoir engineering approach to directionality was recently implemented in optomechanics \cite{fang_generalized_2016}, and also provides a useful way to understand the directional microwave amplifier studied in 
Ref.~\onlinecite{sliwa_reconfigurable_2015} (see also \cite{ranzani_graph-based_2015}).  

In this work, we extend the ideas in \cite{metelmann_nonreciprocal_2015} on two distinct fronts.  The first extension concerns the intuitive underpinnings of our general dissipative scheme.  While in Ref.~\onlinecite{metelmann_nonreciprocal_2015} we described the scheme in terms of reservoir engineering, we show here that in many (but not all) cases, one can understand the scheme as effectively mimicking a perfect measurement plus feed-forward setup, where an observer measures system A, and then continuously uses the results of her measurement to drive system B (cf. Fig.\ref{Fig.:SketchFFScheme}(b)).  We show that the standard continuous quantum measurement theory description of such a system (in the limit of a perfect measurement with vanishing feedback delay) has the same unconditional dynamics as our reservoir engineering protocol.  Given recent work \cite{nelson_experimental_2000, lloyd_coherent_2000, mabuchi_coherent-feedback_2008, james_h^infty_2008, nurdin_coherent_2009,hamerly_advantages_2012,jacobs_coherent_2014}, the fact that there is an intimate connection between our original reservoir-engineering scheme and a measurement based approach is not completely surprising; still, it provides a powerful way to understand the relevant physics that is complementary to the discussion in Ref. \onlinecite{metelmann_nonreciprocal_2015}.

The second new direction explored in this work are the new kinds of behavior possible if one uses {\it two} engineered reservoirs to make a given Hamiltonian interaction between A and B directional.  In particular, we show that this allows one to construct an ``ideal" two-port, phase-preserving bosonic amplifier:  an amplifier that is quantum limited, fully directional, and free from any fundamental gain-bandwidth product limit.  It thus provides a recipe for using two discrete cavity modes plus two engineered reservoirs to mimic the kind of ideal amplification physics that would be exhibited by a perfectly ideal extended traveling wave amplifier.

The remainder of this paper is organized as follows.
We start in Sec.~\ref{sec:Recap} by quickly reviewing the reservoir-engineering approach to non-reciprocity described in Ref.~\cite{metelmann_nonreciprocal_2015}.
In Sec.~\ref{sec:Heuristic}A, we show that for an important class of systems, this reservoir-engineering approach can be rigorously viewed as autonomous feed-forward, with the engineered dissipative reservoir playing the role of an ideal observer and controller.  Using standard continuous measurement theory, we recap the derivation of the unconditional feed-forward master equation for this situation, and show the mapping to our scheme.   In Sec.~\ref{sec:Heuristic}B, we show that in cases where our directional recipe 
cannot be viewed as equivalent to a feed-forward process, it can instead be interpreted as being equivalent to a generalized chiral transport process.  
Finally, in Sec.~\ref{sec:IdealAmp}, we discuss how the use of two engineered reservoirs allows one to construct the ``ideal" bosonic cavity-based amplifier described above.


\section{Recap: non-reciprocity via reservoir engineering}
\label{sec:Recap}

We start by providing a quick recap of the general approach outlined in Ref.~\onlinecite{metelmann_nonreciprocal_2015} for obtaining non-reciprocal interactions between two systems (an approach which is intimately connected to the theory of cascaded quantum systems \cite{Gardiner1993,Carmichael1993,gardiner_quantum_2004}.).  The starting point is two independent systems A and B which interact coherently (and bi-directionally) via a Hamiltonian 
\begin{equation}
	\hat{H}_{\rm int} = \frac{\lambda}{2}\left( \hat{A} \hat{B} + h.c. \right).
	\label{eq:Hint}
\end{equation}
Here, $\hat{A}$ ($\hat{B}$) is a system A (system B) operator; they need not be Hermitian.  We take $[\hat{A},\hat{B}] = 0$, which only imposes an additional restriction in the fermionic case (i.e., in that case $\hat{A}$ and $\hat{B}$ must be built up of even numbers of annihilation and destruction operators).  We also take the interaction strength $\lambda$ to be real without loss of generality.  

To achieve non-reciprocity, we assume both systems have also been jointly coupled to the same engineered Markovian dissipative environment, in such a way that this reservoir mediates the ``dissipative" version of the interaction $\hat{H}_{\rm int}$.  More concretely, the dissipation should be engineered so that the 
full dissipative dynamics of the systems is described by the Lindblad master equation ($\hbar = 1$):
\begin{align}\label{Eq.MasterEqGeneral}
\frac{d}{dt} \hat \rho =& - i \frac{\lambda}{2} \left[ \hat A \hat B + \hat{A}^\dagger \hat{B}^\dagger,\hat \rho \right] + \Gamma \mathcal L \left[ \hat A +  e^{i \varphi} \eta \hat B^\dagger \right] ,
\end{align}  
where the standard dissipative superoperator $\mathcal{L}[\hat{o}]$  is 
\begin{align}\label{Eq.:Superoperator}
	\mathcal L [\hat o ] \hat{\rho} =& \hat o  \hat \rho  \hat o^{\dag} 
					- \frac{1}{2} \hat o^{\dag} \hat o \hat \rho - \frac{1}{2} \hat \rho \hat o^{\dag} \hat o.  
\end{align}
The second term in Eq.~(\ref{Eq.MasterEqGeneral}) describes the effects of the engineered dissipation.  The fact that the dissipation couples to both subsystems implies that it mediates an effective interaction between them (i.e.,~terms $\propto \Gamma \eta$ in Eq.~(\ref{Eq.MasterEqGeneral})).  We refer to this kind of induced interaction as a ``dissipative interaction", as it cannot be described by a Hamiltonian which directly couples systems A and B.  The effects of the dissipation are parameterized by an overall rate $\Gamma$, a dimensionless positive parameter $\eta$ which characterizes the asymmetry of the bath's coupling to the two systems, and a phase $\varphi$.

To show how the master equation (\ref{Eq.MasterEqGeneral}) can be tuned to render directional behavior, 
we consider an arbitrary pair of observables (one for system A, one for system B), described by Hermitian operators $\hat O_A$ and $\hat O_B$ respectively. 
From Eq.~(\ref{Eq.MasterEqGeneral}) we derive the equations of motion for the expectation values: 
\begin{align}
\frac{d}{dt} \ev{\hat O_A } =& - \frac{i}{2} 
	\left[ \lambda + i \Gamma \eta e^{-i \varphi} \right] \ev{\left[ \hat O_A ,\hat A \right] \hat B \hat \rho  }
	\nonumber \\
& - \frac{i}{2} 	\left[ \lambda - i \Gamma \eta e^{+i \varphi}  \right] \ev{\left[ \hat O_A ,\hat A^\dagger \right] \hat B^\dagger \hat \rho  }
	\nonumber \\
                              & + \Gamma \ev{\hat O_A \mathcal L \left[ \hat A \right] \hat \rho} ,
	 \label{eq:OA} \\  
\frac{d}{dt} \ev{\hat O_B } =& - \frac{i}{2} 
	\left[ \lambda - i \Gamma \eta e^{-i \varphi} \right] \ev{\left[ \hat O_B ,\hat B \right] \hat A \hat \rho  }
	\nonumber \\
& - \frac{i}{2} 	\left[ \lambda + i \Gamma \eta e^{+i \varphi}  \right] \ev{\left[ \hat O_B ,\hat B^\dagger \right] \hat A^\dagger \hat \rho  }
	\nonumber \\
                              & + \eta^2 \Gamma \ev{\hat O_B \mathcal L \left[ \hat B^\dagger \right] \hat \rho} .
                              \label{eq:OB}
\end{align} 
The first two terms on the RHS of each equation describe the coupling of the two systems resulting from both the direct (coherent) interaction $\hat{H}_{\rm int}$ and the ``dissipative" bath-mediated interaction.
The last term in each equation describes the additional, purely local dissipative effect of the bath on each subsystem.

It is straightforward to see that we can chose parameters such that the total interaction between the systems becomes directional (e.g.~system A influences the evolution of system B observables, but system A observables evolve independently of system B).  
One needs to balance the amplitude and phase of the dissipative interaction against that of the coherent interaction, i.e.~:
\begin{align}\label{Eq.DirCon}
 \varphi = \pm \frac{\pi}{2}, 
 \hspace{0.5cm}
 \Gamma \eta = \lambda ,
\end{align}
where the sign of $\varphi$ determines the direction of the non-reciprocal interaction (i.e., it defines the direction of the effective information transfer).
Note that these conditions only constrain two of the three dimensionless parameters in our system:  the relative magnitude of the dissipative interaction $\Gamma \eta / \lambda$ and the effective phase of this interaction $\varphi$.  

To be explicit, consider the the case where we want $A$ to influence $B$ but not vice-versa.  This requires tuning the dissipation to satisfy Eqs.~(\ref{Eq.DirCon}) with $\varphi = - \pi/2$.  The EOM for the expectation values then take the form
\begin{align}
	\frac{d}{dt} \ev{\hat O_A } =&  \ \phantom{\eta^2} \Gamma   \ev{\hat O_A \mathcal L \left[ \hat A \right] \hat \rho} ,
 		\label{eq:OAdir} \\  
	\frac{d}{dt} \ev{\hat O_B } =&    
                               - i  \lambda \left[  \ev{ \left[ \hat O_B, \hat B\right]   \hat A \hat \rho }    +   \ev{ \left[ \hat O_B, \hat B^\dagger \right]   \hat A^\dagger  \hat \rho } \right]
                               \nonumber \\
                              &  + \eta^2 \Gamma    \ev{\hat O_B \mathcal L \left[ \hat B^\dagger \right] \hat \rho},  
                        \label{eq:OBdir} 
\end{align}
with $\lambda = \eta \Gamma$. 
System B is ``forced" by system A in the same way it would be if we had no dissipation and only the coupling $\hat{H}_{\rm int}$, i.e.,~compare the order-$\lambda$ terms in 
Eq.~(\ref{eq:OB}) with those in Eq.~(\ref{eq:OBdir}).  In contrast, system A observables evolve in a manner independent of system B; they only feel a local dissipative effect from the coupling to the reservoir.

Finally, we note that for non-Hermitian case, one could equally as well obtained directionality using a master equation
\begin{align}\label{Eq.MasterEqAlt}
	\frac{d}{dt} \hat \rho =& - i \frac{\lambda}{2} \left[ \hat A \hat B + \hat{A}^\dagger \hat{B}^\dagger,\hat \rho \right] + 
	\Gamma \mathcal L \left[ \hat A^\dagger -  e^{-i \varphi} \eta \hat B \right] ,
\end{align}
Tuning parameters as per Eq.~(\ref{Eq.DirCon}) again gives rise to directional interactions.  While the coupling terms in this scheme are identical, the local damping terms generated by the dissipation will be different:  in the last term of Eq.~(\ref{eq:OA}), $\hat{A}$ will be replaced by $\hat{A}^\dagger$, and in the last term of Eq.~(\ref{eq:OB}), $\hat{B}^\dagger$ will be replaced by $\hat{B}$.  
.

\section{Heuristic insight into dissipative non-reciprocity}
\label{sec:Heuristic}


\subsection{Mapping to a measurement based feed-forward protocol}

While the algebra underlying our general dissipation-based recipe for non-reciprocity is straightforward enough, the underlying intuition for {\it why} such a scheme works
may remain opaque.  Here, we demonstrate that for the case where the coupling operators $\hat{A}$ and $\hat{B}$ in Eq.~(\ref{eq:Hint}) are Hermitian, there is a direct correspondence between our reservoir-engineering approach and the situation sketched in Fig.~\ref{Fig.:SketchFFScheme}b.  This figure depicts a simple way to generate a non-reciprocal interaction between systems A and B:  an observer makes a continuous measurement of the system $A$ observable $\hat{A}$, and uses her measurement record to then apply an appropriate force to system $B$; this force is taken to couple to the observable $\hat{B}$.  Even at a heuristic level, one could imagine that in such a feed-forward setup, the net result would be a directional interaction of the form we are after.  We now show rigorously that in the limit where both the measurement and forcing steps are ideally performed, the unconditional dynamics of this feed-forward setup results in {\it exactly} our directional master equation defined by Eqs.(\ref{Eq.MasterEqGeneral}) and (\ref{Eq.DirCon}).  This demonstrates that at least for Hermitian coupling operators $\hat{A}$ and $\hat{B}$, our scheme is the reservoir-engineering (or coherent feedback) equivalent of the measurement based protocol.

Our derivation makes use of standard continuous quantum measurement theory (see, e.g., Ref.~\onlinecite{Jacobs2006,wiseman_quantum_2010} for recent pedagogical reviews).  We start by describing the continuous monitoring of the system A observable $\hat{A}$ by our observer.  The measurement record in a particular run of the experiment $I(t)$ is determined by:
\begin{align}\label{Eq.:MeasurmentRecord}
   dI(t) =   \sqrt{k}  \ev{ \hat A (t)} dt   + dW(t),
\end{align}
where $k$ is a rate representing the strength of the measurement, and  
$dW$ is a standard Wiener increment (describing the white imprecision noise in the measurement record).  It fulfills
the usual conditions $\overline{dW} = 0$ and $\overline{dW^2} = dt$, where the average here is over measurement outcomes.  
The expectation value of $\hat{A}$ appearing above is determined by the conditional density matrix of system, $\hat{\rho}_c$,  which evolves as:
\begin{align}\label{Eq.:MasterEqCondSt}
d \hat \rho_c =&   \frac{k }{4} \mathcal L \left[\hat A  \right] \hat \rho_c dt
		+  \frac{\sqrt{k}}{2}  \left[\hat A \hat \rho_c + \hat \rho_c \hat A -  2 \ev{\hat A} \hat \rho_c \right] dW . 
\end{align} 

Next, we describe our observer's forcing of system B using the measurement record $I(t)$; we assume the observer forces system B by explicitly coupling to the Hermitian system-B operator $\hat{B}$.   As a result of this forcing, system B will thus evolve under the Hamiltonian
\begin{align}
 	\hH_{\rm FF}(t) = \sqrt{\alpha_{\rm FF}} I(t-\tau) \hat B,
\end{align}
where $\alpha_{\rm FF}$ corresponds to the strength of the applied forcing and has the units of a rate.
We now want to write the full evolution of the density matrix system A and B, averaged over all possible measurement outcomes (i.e.,~the unconditional master equation). 
One needs to expand the master equation to second order before taking the average over outcomes; see Ref.~\onlinecite{wiseman_quantum_2010} for a detailed derivation.  Taking the limit $\tau \rightarrow 0^+$ (i.e.,~negligible delay in applying the feed-forward force), one obtains the standard feed-forward (or feedback) master equation (FME) \cite{caves_quantum-mechanical_1987, wiseman_quantum_1994, doherty_quantum_2000, wiseman_quantum_2010}   
\begin{align}\label{Eq.:MasterEqFB}
\frac{d}{dt} \hat \rho =&        \frac{k }{4} \mathcal L  \left[\hat A  \right] \hat \rho 
				+ \alpha_{\rm FF} \mathcal L       \left[\hat B \right] \hat \rho 
		- i   \frac{\sqrt{k \ \alpha_{\rm FF}}}{2}    \left[\hat B, \hat A\hat \rho  + \hat \rho  \hat A     \right]  .
\end{align}
One can show that this equation is indeed in Lindblad form.  The first term describes the backaction disturbance of system A due the measurement, while the second term accounts for the fact that noise in the feed-forward force (due to noise in the measurement record) causes a dissipative evolution of system B.  Only the last term describes the effective directional coupling between the two systems induced by the feed-forward protocol.  

While it may not be obvious, this master equation does indeed describe a completely non-reciprocal interaction between systems A and B, where system A is completely uninfluenced by the evolution of system B (but not vice-versa).  A few lines of algebra show that it is completely equivalent to our directional master equation (i.e.,~Eq.~(\ref{Eq.MasterEqGeneral}) with the constraint of Eq.~(\ref{Eq.DirCon}))  if one makes the mapping:
\begin{eqnarray}
	\alpha_{\rm FF}  & = \eta^2  \Gamma ,\\
	 k   & = 4\Gamma .
\end{eqnarray}

We have thus demonstrated that for Hermitian coupling operators $\hat{A}$ and $\hat{B}$, our reservoir-engineering protocol is formally equivalent to the measurement-based feed-forward protocol.  Despite this equivalence, our approach nonetheless possesses many practical advantages.  The above derivation of the feed-forward master equation assumed a perfectly quantum limited measurement of $\hat{A}$, as well as a perfect implementation of the feed-forward force.  These conditions would obviously be difficult to implement in practice.  In contrast, the reservoir engineering approach does not require a quantum limited measurement or any ideal processing of a classical measurement record.


\subsection{Mapping to interactions via chiral transport}

An alternate method for obtaining some intuition into the form of Eq.~(\ref{Eq.MasterEqGeneral}) is to show how it would arise if one could couple system A and B to an explicitly 
one-dimensional chiral bosonic waveguide.  The directionality inherent in this master equation is then seen to directly mirror the directional bosonic transport in the waveguide.  Unlike the measurement-plus-feed-forward derivation of the previous section, this ``chiral transport" derivation also applies if the coupling operators $\hat{A}$ and $\hat{B}$ are non-Hermitian.  The derivation here mirrors exactly the original derivation of cascaded quantum systems given by Carmichael \cite{Carmichael1993} (see also Ref.~\onlinecite{gardiner_quantum_2004}).  

Following Ref.~\onlinecite{Carmichael1993}, we start by assuming that systems A and B couple locally to a zero-temperature chiral 1D bosonic waveguide of right-moving photons (field operator $\hat{\psi}(x)$) at positions $x=0$ and $x=d$ ($d >0$).  This is described by an interaction Hamiltonian of the form:
\begin{equation}
	\hat{H}_{\rm SB} =
		i \sqrt{\kappa_A} \left( \psi^\dagger(0) \hat{A} - h.c. \right) +
		i \sqrt{\kappa_B} \left( \psi^\dagger(d) \hat{B}^\dagger - h.c.\right ) .
\end{equation}
Here, $\kappa_A$ and $\kappa_B$ parameterize the strengths of the couplings.   As is demonstrated in Ref.~\onlinecite{Carmichael1993} by making a standard Born-Markov approximation, 
one can derive a master equation for the ``source retarded" reduced density matrix $\hat{\rho}_R(t)$ describing the state of system B at time $t$ and system A at time $t = t - d/ v$, where $v$ is the waveguide velocity.  In the limit where $d \rightarrow 0^+$, the time shift $d / v$ can be neglected, and this master equation is the generalized version of the standard cascaded quantum systems master equation.  If we make the gauge change $\hat{B} \rightarrow i \hat{B}$, one can easily confirm that it has
the form of our general master equation Eq.~(\ref{Eq.MasterEqGeneral}) with the correspondence:
\begin{equation}
	\lambda  = \sqrt{\kappa_A \kappa_B} , \,\,\,
	\Gamma  = \kappa_A , \,\,\,
	 \eta  = \sqrt{\kappa_B / \kappa_A}  ,  \,\,\,
	 \varphi  = - \pi/2.
\end{equation} 
Comparison against Eq.~(\ref{Eq.DirCon}) shows that the conditions needed for directionality (i.e.,~balancing of coherent and dissipative interactions) are of course satisfied.  We thus see that the parameters in our effective directional master equation can be tied to the strength of the two couplings to the chiral waveguide.

\begin{figure} 
  \centering\includegraphics[width=0.4\textwidth]{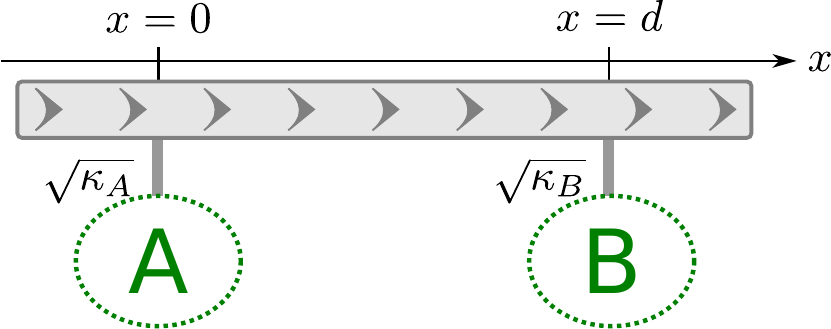} 
 	\caption{Coupling system A and B to a 1D chiral waveguide with coupling strength $\sqrt{\kappa_{A}}$ and  $\sqrt{\kappa_{A}}$ respectively.  The coupling is such that a photon can be created in the waveguide 
 	        at position $x=0$ by acting on system $A$ with the operator $\hat{A}$. This photon then propagates to $x=d$, where it can be destroyed while simultaneously having the operator $\hat{B}$ act on system $B$.  The effective dynamics is then equivalent to the general dissipation-based directionality recipe.
 	\label{Fig.:SketchCascaded}}
\end{figure}

The above derivation gives some intuition for how directional interactions arise.  At $x=0$, the interaction Hamiltonian $\hat{H}_{\rm SB}$ can create a photon in the waveguide while simultaneously acting on system A with operator $\hat{A}$.  This excitation then propagates to the right in the waveguide.  When it reaches $x=d$, $\hat{H}_{\rm SB}$ can destroy this photon while simultaneously acting on system B with the operator $\hat{B}$.  We thus have the needed directional interaction:  the action of $\hat{A}$ followed by $\hat{B}^\dagger$.  The converse process is of course impossible as there is only rightwards propagation in the waveguide.  

While the above model with a chiral waveguide helps provide intuition into our general directional master equation, it is not intended as being the preferred recipe for how to implement directionality.  As discussed in detail in Ref.~\onlinecite{metelmann_nonreciprocal_2015} , one can in many cases implement the directional version of our master equation in Eq.~(\ref{Eq.MasterEqGeneral}) by simply using parametric couplings between localized cavity modes.  Remarkably, these systems are able to completely mimic the chiral transport situation described above:  in both cases, the final dynamics of the relevant system modes are directional in the same manner.  We also stress that the above derivation applies to both Hermitian and non-Hermitian coupling operators $\hat{A}, \hat{B}$.  In the Hermitian case, we can view the chiral waveguide as the observer performing the feed-forward operation described in the previous section.  

Finally, it is worth noting that for non-Hermitian coupling operators, the direct interaction $\hat{H}_{\rm int}$ could always be expressed as the sum of two terms, each being the product of only Hermitian operators; one simply uses the Hermitian and anti-Hermitian parts of $\hat{A},\hat{B}$.  For each one of these terms, one could then use our directionality recipe for Hermitian coupling operators.  The result would be a master equation having two dissipators.  It would be completely equivalent to two measurement-plus-feedforward processes.  We stress that this is {\it not} the same as directly implementing the directional version of Eq.~(\ref{Eq.MasterEqGeneral}), something that only requires a single reservoir.  In this case, one cannot interpret the final master equation using the measurement-plus-feedforward interpretation of Sec.~\ref{sec:Heuristic}A; only the ``chiral transport" interpretation of this section holds.  This difference can have important physical consequences.  It implies that for Hermitian coupling operators, our directional master equation can never generate entanglement (as it is equivalent to performing local operations and classical communication).  For non-Hermitian coupling operators, this is no longer true, and the directional master equation can in principle generate entanglement between systems A and B.


\section{The \textit{ideal} quantum amplifier}
\label{sec:IdealAmp}

A highly desired component in many quantum measurement protocols is a directional quantum amplifier.
It realizes non-reciprocal amplification of weak input signals, implying that any noise
generated at the amplifier output (by e.g.~more classical amplifiers higher up in the measurement chain) does not get amplified and drives the typically fragile quantum signal source.  Directional amplifiers
can reduce the number of circulators needed in a measurement setup; given that such circulators are often lossy and bulky, this is a great practical advantage.  
So far all experimental realization for directional amplifiers are based on a superconducting circuits architecture, e.g.
traveling wave amplifiers \cite{obrien_resonant_2014,white_traveling_2015} or Josephson parametric converters setups \cite{kamal_gain_2012,abdo_directional_2013, abdo_josephson_2014,sliwa_reconfigurable_2015}.

\begin{figure} 
  \centering\includegraphics[width=0.45\textwidth]{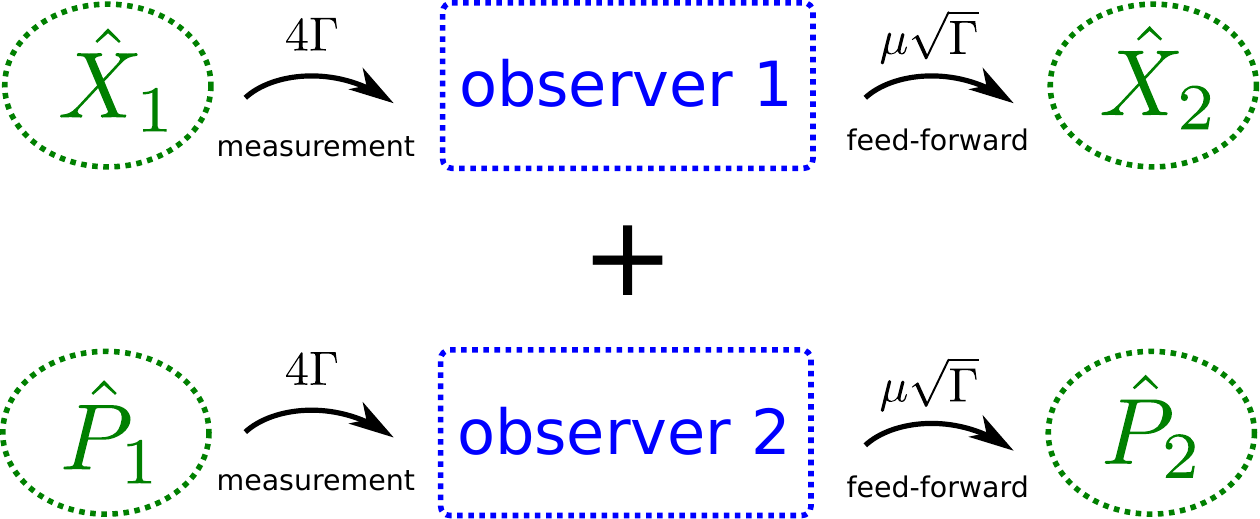} 
 	\caption{Illustration of the \textit{ideal} amplifier scheme involving two measurement feed-forward schemes.
 	Observer 1 and 2 measure both cavity-1 quadratures and feed-forward the measurement records to cavity-2, i.e.,
 	the whole information of the quadratures $\hat X_1$ and $\hat P_1$ is transferred to $\hat X_2$ and $\hat P_2$ but not vice versa.  
	}
 	\label{Fig.:DPAscheme}
\end{figure}

In this section, we show how to construct a truly \textit{ideal} directional quantum amplifier.
By ideal we mean more than just saying that the amplifier has quantum-limited noise:  we also want the amplifier to be phase preserving, i.e., both quadratures of the photonic field are amplified,
and to have an amplification bandwidth that does not degrade with increasing gain (i.e.,~it is not limited by a conventional gain-bandwidth constraint). 
As discussed in our previous work \cite{metelmann_nonreciprocal_2015}, either one of these properties (but not both simultaneously) could be achieved in a directional amplifier by using a single engineered reservoir.
We show now that one can achieve both of these desirable conditions in a design that utilizes two engineered reservoirs.
The setup basically corresponds to the combination of two feed-forward protocols, as illustrated in Fig.~\ref{Fig.:DPAscheme}. 
While a direct implementation of these feed-forward processes would suggest a rather complicated coupling to the engineered reservoirs, we show that in fact a relatively simple implementation is possible.


\subsection{A directional phase in-sensitive amplifier via two engineered reservoirs}

We start with a setup of two cavity modes (lowering operators $\hat{d}_1$, $\hat{d}_2$) described via the quadrature field operators $\hat X_{n},\hat P_{n} (n \in 1,2)$, i.e.,~$\hat{d}_n = (\hat{X}_n + i \hat{P}_n) / \sqrt{2}$.
We take the direct, coherent interaction between the modes to take the form
\begin{align}\label{Eq.HcohNDPA}
 \hH_{\COH}^{\pm}  = J \left(  \hat X_1 \hat X_2  \pm \hat P_1  \hat P_2  \right).
\end{align}
One can easily check that the $-$ sign realizes a phase-insensitive parametric amplifier interaction between the modes, while the $+$ sign corresponds to a simple hopping or beam splitter interaction.  
We work in a rotating frame where the two cavities are effectively resonant, and thus the Hamiltonian is time independent.

To proceed, note first that if we only kept half of the interaction in Eq.~(\ref{Eq.HcohNDPA}) 
(e.g., $  \hH_{\rm coh}  = J \hat X_1 \hat X_2 $), the resulting QND interaction would realize a phase sensitive coherent amplifier 
with no gain-bandwidth limitation (see Sec. IIC of \cite{metelmann_nonreciprocal_2015}). To render 
such an interaction directional using our general recipe, we would have to combine it with a dissipative interaction described 
by a superoperator of the form $\mathcal L  [  \hat X_1  - i \hat X_2 ]$ \cite{metelmann_nonreciprocal_2015}.  
This results in a directional interaction from cavity 1 to cavity 2 involving only a single quadrature, 
i.e.,~ the phase information is lost.

Our goal here however is to have a directional interaction where phase information is not lost, i.e.,~both quadratures of cavity 1 have to directionally force cavity 2.  At the same time, we want to keep the absence of a gain-bandwidth limit on our amplifier.  Our approach is simple: we will apply our general recipe twice, introducing one engineered reservoir to make the $\hat{X}_1 \hat{X}_2$ interaction directional, and another to make the $\hat{P}_1 \hat{P}_2$ interaction directional.  We thus need two nonlocal dissipators.
The corresponding master equation has the form
\begin{align}\label{Eq.MasterEqNDPA}
 \frac{d}{dt} \hat \rho = & - i \left[\hH_{\COH}^{\pm}, \hat \rho \right] 
		+ \Gamma \mathcal L \left[  \hat X_1  -   i \eta \hat X_2  \right] \hat \rho
		+ \Gamma \mathcal L \left[  \hat P_1 \mp  i \eta \hat P_2  \right] \hat \rho. 
\end{align} 
Note that we have taken the coupling rate $\Gamma$ and asymmetry parameter $\eta$ characterizing the
coupling to the engineered reservoirs to be the same for each of the two reservoirs; this ensures that ultimately, both input signal quadratures incident on cavity $1$ will be amplified equally. 
As per our recipe, we have also picked the relative phase between the two terms in each jump operator (i.e.,~$\pm i$) to make both interactions directional, in that cavity 1 is not influenced by cavity 2.

From the master equation (\ref{Eq.MasterEqNDPA}) we can derive the equations of motion for the cavity modes expectation values, they read
\begin{align}\label{Eq.:EoMquadratures}
 \frac{d}{dt}  \ev{\hat X_1} 
		=& \pm \left[ J - \eta \Gamma \right] \ev{ \hat P_2 },
\nonumber \\ 
 \frac{d}{dt}  \ev{\hat P_1} 
		=&  -  \left[ J - \eta \Gamma \right] \ev{ \hat X_2  },
\nonumber \\
 \frac{d}{dt}  \ev{\hat X_2} 
		=& \pm \left[ J + \eta \Gamma \right] \ev{\hat P_1 },
\nonumber \\
 \frac{d}{dt} \ev{\hat P_2} 
		=&   - \left[ J + \eta \Gamma \right] \ev{\hat X_1}. 
\end{align} 
By setting $J = \eta \Gamma $ in Eq.~(\ref{Eq.:EoMquadratures}) the system becomes directional as desired: cavity 2 is influenced by cavity 1, but not vice-versa.

We use the directional interaction between the two cavities to construct an directional quantum amplifier, where an input signal on cavity 1 leaves cavity 2 amplified and
with not more added noise as allowed by quantum mechanics \cite{caves_quantum_1982}. To understand how signals are transferred through the system we move away from
the master equation description and utilize quantum Langevin equations.
We couple both cavities to input output waveguides with coupling strength $\kappa$ and model the engineered reservoir as a Markovian oscillator bath.
This allows us to use input/output theory as usual, and derive the scattering matrix relating outputs in
the two waveguides to corresponding inputs.  
We work in the basis $\mathbf{\hat D}  = [ \hat X_1 , \hat P_1 , \hat X_2 ,\hat P_2 ]^T$ and obtain the scattering matrix $  \mathbf{s}  $ on resonance (zero frequency in this rotated frame)
\begin{align} \label{Eq.:ScattMat}
  \mathbf{s}[\omega=0]  &= 
\left(
\begin{array}{cccc}
	- 1     & 0            &   0       &   0
\\
	  0     &- 1  	       &   0       &   0 
\\
	 0       & \mp\sqrt{\mathcal G_0}       & - 1   &  0            
\\
	\sqrt{\mathcal G_0} & 0              & 0  &   - 1             
\\
\end{array}
\right),  
\hspace{0.2cm}
\mathbf{\hat D}_{\rm out} =   \mathbf{s}  \mathbf{\hat D}_{\rm in} + \vec{\xi},
\end{align}
where $\sqrt{\mathcal G_0}  = \frac{8 J}{ \kappa}$ corresponds to the amplitude gain.
Crucially, we have to take into account that there will be additional noise ($\vec{\xi}$) originating from the engineered reservoirs. Without these noise contributions, one would erroneously conclude that our
system (a phase-preserving amplifier) violates the quantum limit on added noise. We discuss the noise properties in more detail in Sec.~\ref{sec:IdealAmp}C, showing that the added noise (including all contributions) is at the minimal level required by the quantum limit.

The scattering matrix Eq.(\ref{Eq.:ScattMat}) describes directional phase-insensitive amplification, the cavity-1 input-quadratures $\hat X_{1, \rm in}$ an $\hat P_{1, \rm in}$ leave cavity 2 amplified,
while any input on cavity 2 will never show up at cavity 1.  Note that the there is a unity
reflection of signals and noise incident on each cavity (described by the diagonal elements of $\mathbf{s}$).  We stress that there is no gain associated with these reflections.  The unity level of reflection could in principle be suppressed using impedance matching techniques (see Ref.\cite{metelmann_nonreciprocal_2015}). 

\begin{figure*} 
  \centering\includegraphics[width=1.0\textwidth]{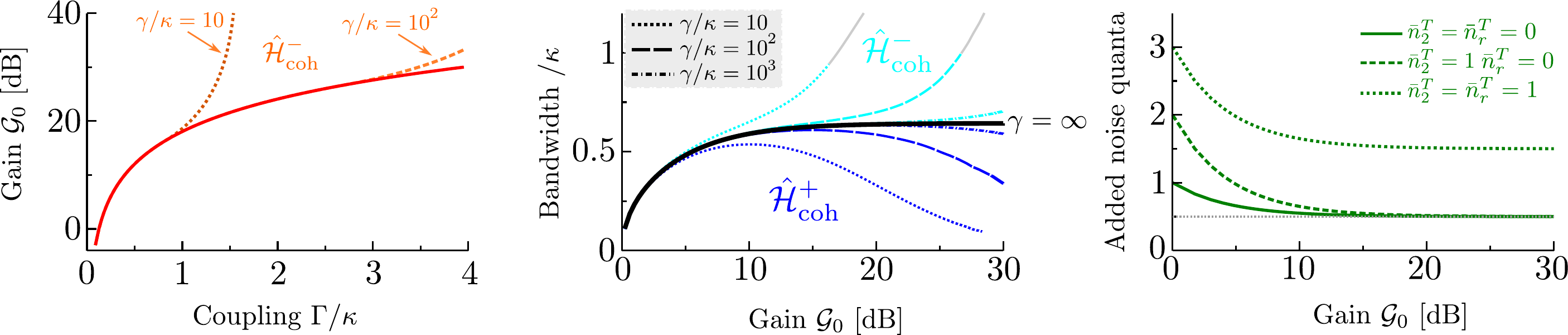} 
 	\caption{Characteristics of the directional phase-insensitive quantum limited amplifier. The left plot depicts the gain as a function of the coupling strength $\Gamma$.
 	The red solid line corresponds to the case of a perfect Markovian reservoir, i.e., $\gamma/\kappa \rightarrow \infty$, where the zero-frequency gain scales as $\mathcal G_0 = (8 \Gamma/\kappa)^2$.
 	The orange dashed and dotted lines correspond to the case of finite ratios of the decay rates $\gamma/\kappa$ for $\mathcal H_{\rm coh}^{-} $, parameters as denoted in the graph.  
 	There we observe mode-splitting, i.e., the gain curve has a double peak structure and the maxima of the peaks increases strongly with $\Gamma/\kappa$. 
 	The middle graph shows results for the bandwidth. Including non-Markovian effects the bandwidth decreases for a coherent beam-splitter interaction $\mathcal H_{\rm coh}^{+}$, while it increases for the coherent amplifying interaction $\mathcal H_{\rm coh}^{-}$, until the mode-splitting regime is reached (gray solid lines).
 	The right graph depicts the added noise for the system, assuming $\bar n_{r,1}^{T} = \bar n_{r,1}^{T} \equiv \bar n_{r}^{T} $.
	}
 	\label{Fig.:AmplifierProperties}
\end{figure*}


\subsection{Simpler coupling to dissipation}

Each dissipator appearing in Eq.~(\ref{Eq.MasterEqNDPA}) could in principle be realized by having a nonlinear (i.e.,~parametric) coupling between cavities $1$ and $2$ and an auxiliary, highly
damped mode that plays the role of a bath.  By applying appropriate coherent pump drives, the required dissipator can be obtained; this strategy is discussed in great detail in Ref.~\cite{metelmann_nonreciprocal_2015}, and was even implemented experimentally in Ref.~\cite{fang_generalized_2016}.  The pump tones and nonlinearity basically allow one to couple the process of creating or destroying a photon in a given cavity to creating a quanta in the dissipative mode (which then rapidly dissipates).

More formally,
let $\hat{c}$ denote the lowering operator of the heavily-damped bath mode that will play the role of one of our engineered reservoirs.  If this mode has nonlinear interactions with the cavity modes, then the application of strong pump tones can result in a 
mean-field interaction Hamiltonian of the form
\begin{align}
	\hH_{\rm SB} = \lambda \hat c^{\dag} \hat z + h.c. ,
\end{align}
where $\hat {z}$ is a linear combination of raising and lowering operators for both cavity 1 and 2, and $\lambda$ is an effective interaction amplitude.  In
the limit where $\hat{c}$ is strongly damped, it can be eliminated from the dynamics, resulting in a Lindblad
superoperator of the form
\begin{align} 
 \frac{4 |\lambda|^2}{\gamma} \mathcal L \left[ \hat z\right] .
\end{align}
Here $\gamma$ denotes the decay rate of the auxiliary mode.  This provides one route for generating the dissipators needed for our ideal directional quantum amplifier setup.

This being said, the kind of dissipators required in Eq.~(\ref{Eq.MasterEqNDPA}) involve jump operator like $\hat{X}_1 + i \hat{X}_2$.  These involve photon creation and destruction in both cavity $1$ and $2$, and would in principle require four distinct pump tones.  On the surface, this would make realizing 
Eq.~(\ref{Eq.MasterEqNDPA}) quite challenging, as 8 pump tones would be necessary.

Remarkably, a much simpler strategy is possible requiring only 4 pump tones.  The dissipators required in Eq.~(\ref{Eq.MasterEqNDPA}) can be replaced by two far simpler dissipators (for both cases of sign):
\begin{align}\label{Eq.QuadDbasis1}
     \mathcal L \left[  \hat X_1 - i \hat X_2  \right] +   \mathcal L \left[  \hat P_1  - i \hat P_2 \right]
=&   \mathcal L \left[  \hat d_1 - i \hat d_2  \right] +   \mathcal L \left[ \hat d_1^{\dag}  - i \hat d_2^{\dag}  \right],
\end{align}
\begin{align}\label{Eq.QuadDbasis2}  
    \mathcal L  \left[  \hat X_1 - i \hat X_2  \right] +   \mathcal L \left[  \hat P_1  + i \hat P_2 \right]
=&  \mathcal L  \left[  \hat d_1 - i \hat d_2^{\dag} \right] +   \mathcal L \left[   \hat d_1^{\dag} - i \hat d_2 \right].
\end{align}
While mathematically equivalent, the RHS of the equations above represent a simpler method for practically implementing our scheme.  An explicit method for implementing the scheme in optomechanics (using the simplified dissipators) is presented in Appendix~\ref{App:OMrealization}


\subsection{Added noise and finite frequency gain} \label{Sec.:NoiseFrequencyGain}

While we have established that our two-reservoir amplifier scheme is indeed directional and phase preserving,
we have not established two other crucial desired properties:  the absence of a standard gain-bandwidth product, and the presence of quantum-limited added noise.  We address both properties in this subsection.  

Let us address first the question of the  bandwidth over which input signals can be amplified. 
We still consider the situation were the forward direction is from cavity 1 and 2, and the reverse direction, from cavity 2 to 1, is blocked.
Standard input-output theory yields for the frequency-dependent forward photon-number gain
\begin{align}
 \mathcal G[\omega] \equiv \left|s_{32}[\omega]\right|^2 = \left|s_{41}[\omega]\right|^2  =&  \frac{ \mathcal G_0  }
                             { \left[1 + \frac{4 \omega^2}{\kappa^2} \right]^2  }, 
\end{align}
with the zero-frequency gain $\mathcal G_0 $ as defined after Eq.~(\ref{Eq.:ScattMat}). 
The gain is simply a Lorentzian squared with bandwidth in the order of $\kappa$. Crucially, the bandwidth does not depend on the gain, thus we have 
no fixed gain-bandwidth limit.

We also want to make sure that our amplifier remains directional over the full amplification bandwidth.  Again, standard input-output theory yields for the reverse gain:  
\begin{align}
 \bar{\mathcal G}[\omega] \equiv \left|s_{23}[\omega]\right|^2 = \left|s_{14}[\omega]\right|^2 = 0.
\end{align}
As desired, the reverse gain vanishes completely even for finite frequency.
We have of course presented calculations for the limit where the engineered dissipation has a vanishingly small correlation time, i.e.,~in the Markovian limit.  For results involving a finite memory time of the engineered reservoir see Appendix~\ref{App:NonMarkovian}.

The remaining question concerns the added noise of our amplifier and its magnitude compared to the quantum limit value.  From input-output theory, we find that the  
the added noise of the amplifier at zero frequency (expressed as an effective number of quanta at the amplifier input) is given by
\begin{align}\label{Eq.:AddedNoiseQuadratureBasis}
 \bar n_{2, \rm add}  =&  \frac{1}{2}   
			 +  \frac{1}{2}     \left( \bar n_{r,1}^T  + \bar n_{r,2}^T   \right) 
			 +  \frac{1}{\mathcal G_0} \left(\bar n_2^T + \frac{1}{2} \right),
\end{align}
where we set $ \Gamma   =  \frac{\kappa}{2}  $ and $ \eta = 2 \sqrt{\mathcal G_0} $ for optimal noise performance.
Here $\bar n_{r,n}^T $ correspond to the averaged thermal occupancies of the engineered baths, while $\bar n_2^T$ characterizes the thermal noise incident on cavity 2.  In the large gain limit,
we see that the quantum limit is indeed achieved as long as the two engineered reservoirs are at zero temperature, i.e., $\bar n_{r,1}^T = \bar n_{r,2}^T = 0$.  Note that small thermal occupancies of the reservoirs lead to only a small deviation from the quantum limit.

Finally, note that the added noise in Eq.~(\ref{Eq.:AddedNoiseQuadratureBasis}) was calculated 
for Markovian reservoirs corresponding to the two dissipators given in Eq.~(\ref{Eq.MasterEqNDPA}).  If one instead realizes our ideal amplifier using the simpler dissipators given on the RHS of Eqs.~(\ref{Eq.QuadDbasis1})
or Eqs.~(\ref{Eq.QuadDbasis2}), the added noise is
given by Eq.~(\ref{eq:AddedNoiseDBasis}).  It is identical to the expression in Eq.~(\ref{Eq.:AddedNoiseQuadratureBasis}) as long as the two reservoirs have identical thermal occupancies (i.e.~$\bar n_{r,1}^T = \bar n_{r,2}^T$).


\section{Conclusion}

In this article, we have extended the general recipe for directional interactions presented in 
our previous publication Ref.~\cite{metelmann_nonreciprocal_2015}.  
We demonstrated that in many cases, this general recipe is equivalent
to the dynamics that would result from a continuous measurement based feed-forward protocol. The latter is directional from its basic nature and provides crucial intuition into the underlying mechanism
of our recipe.  We also extended the application of our ideas to the field of directional quantum amplifiers, showing the utility of using two engineered reservoirs.  This allows for the construction of a quantum-limited directional amplifier that is both phase-preserving, and which does not suffer from a standard gain-bandwidth constraint.   


\acknowledgments{
This work was supported by the DARPA ORCHID program through a grant from AFOSR, by the University of Chicago quantum engineering program, and the AFOSR MURI program.}


\appendix


\section{Non-Markovian effects for finite frequency }\label{App:NonMarkovian}

In the main text we considered the Markovian limit when deriving the gain and reverse gain for our ideal directional amplifier.  In this appendix, we consider
consider the more general case where the engineered reservoirs have a finite correlation time and thus deviate from the Markovian limit

We start by modeling the reservoirs
with the system-bath Hamiltonian  
\begin{align}\label{Eq.:SystemBathNDPA}
 \hH_{\rm SB}^{\pm} =   \lambda_1 \left[   \hat X_1 \hat V_1   +   \hat P_1 \hat V_2 \right]  
		+ \lambda_2 \left[   \hat X_2 \hat U_1  \pm  \hat P_2 \hat U_2 \right].
\end{align}
Here, we have two auxiliary damped modes that act as the non-Markovian reservoirs; they are described by  the quadrature operators $\hat U_{n}$ and $\hat V_{n}$, $(n =1,2)$. Both auxiliary modes are themselves coupled to Markovian reservoirs, associated with  the decay rates $\gamma_{n}\equiv \gamma$.  The effective correlation time of the reservoirs is given by $1 / \gamma$. We combine the interaction $\hH_{\rm SB}^{\pm}$ with the coherent interaction $\hH_{\COH}^{\pm}$ in Eq.~(\ref{Eq.HcohNDPA}).
The condition to achieve directionality now becomes $ J = \frac{2\lambda_1 \lambda_2  }{\gamma }$. 

Again we can use input/output theory to calculate the scattering matrix for the whole 6 mode system.
We consider the situation where we aim for directional amplification from cavity 1 to cavity 2.
The gain and reverse gain for finite frequency  become
\begin{align}\label{Eq.GainNonMark}
 \mathcal G^{\pm}[\omega] =&        
                 \frac{ \mathcal G_0 \left[ 1 + \frac{ \omega^2 }{\gamma^2} \right] \left[ 1 + \frac{4\omega^2 }{\gamma^2}\right]}
                      {\left| \left[   1 - i \frac{2\omega }{\gamma}  \right]^2 \left[1 - i \frac{2 \omega}{\kappa} \right]^2  \mp \frac{i}{4}    \frac{ \omega }{\gamma}   \mathcal G_0    
                        \left[ 1- i \frac{ \omega }{\gamma}     \right]    \right|^2},
                        \nonumber \\  
 \bar{ \mathcal G}^{\pm}[\omega]  =&    \frac{\frac{\omega^2}{\gamma^2} }{\left[ 1 + \frac{ \omega^2 }{\gamma^2} \right]}  \mathcal G^{\pm}[\omega],
\end{align}
as expected for zero frequency the reverse gain vanishes and the gain takes the value $\mathcal G_0$.
Note that the above expressions do not assume anything about the magnitude of the decay rate $\gamma$ of the auxiliary reservoirs.  We see that the directionality on resonance is a robust feature and does not require
Markovian limit where $\gamma$ is extremely large.  Having a large $\gamma$ is however crucial if one wants directionality over a large bandwidth.   The reverse gain is suppressed for frequencies $\omega \ll \gamma$,
hence, for large $\gamma$ we have directionality over a large frequency regime.

Another important point is that the denominator of the gain
$\mathcal{G}^{\pm}[\omega]$ contains a term with the zero-amplitude gain squared, which is suppressed in the strong damping regime. This term can lead to mode-splitting for the case of a coherent parametric amplifier interaction between the two principal cavity modes, i.e., for $\mathcal G^{-}[\omega]$ in  Eq.~(\ref{Eq.GainNonMark}).
Moreover, for a small ratio $\gamma/\kappa$ the amplification bandwidth is affected, i.e., the bandwidth increases (decreases) for $\hH_{\COH}^{-}(\hH_{\COH}^{+})$, cf. middle graph of Fig.~\ref{Fig.:AmplifierProperties}.
We do not analyze this in more detail here.  In general, one sees that having a large $\gamma$ is favorable.


\section{An optomechanical realization of the \textit{ideal} amplifier}\label{App:OMrealization}

\begin{figure} 
  \centering\includegraphics[width=0.4\textwidth]{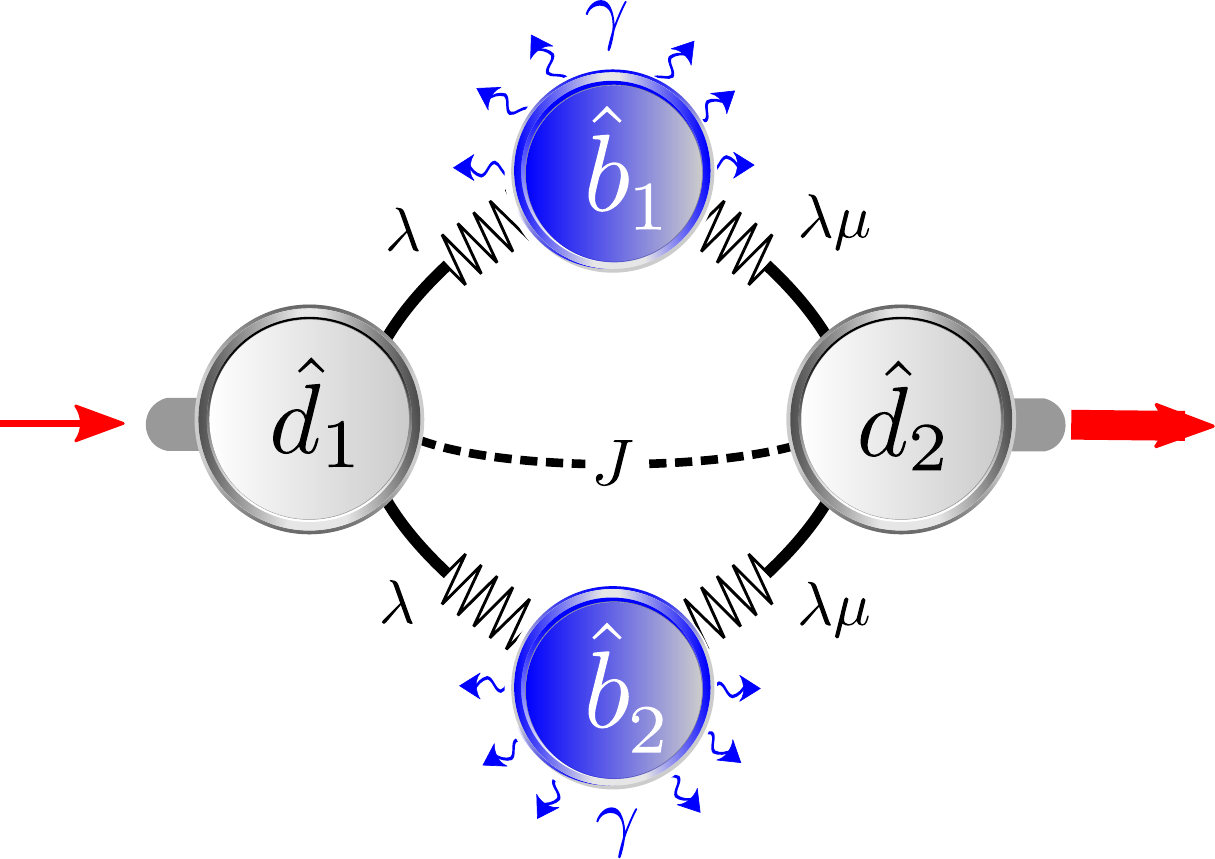} 
 	\caption{Sketch of the optomechanical realization of a quantum-limited, phase-insensitive and directional amplifier without a gain bandwidth product.
 	        Two cavity modes ($\hat d_1$ and   $\hat d_2$) are coupled directly via a coherent hopping interaction of strength $J$, and indirectly
 	        via two mechanical modes ($\hat b_1$ and   $\hat b_2$). The Markovian limit is reached if the mechanical modes are strongly damped, i.e., for large $\gamma$,
 	        in that case the mechanical modes can be adiabatically eliminated and the effective master equation description holds, cf. Eq.~\ref{Eq.MasterEqNDPA}.}
 	\label{Fig.:SketchOMsetup}
\end{figure}

The two reservoir phase-insensitive amplifier introduced in the main text could for example be realized in an optomechanical setting, cf.  Fig.\ref{Fig.:SketchOMsetup}. 
Here two mechanical modes are used to realize the required engineered reservoirs.
We work in the basis of non-Hermitian operators, where $\hat b_{1,2}$ denote the mechanical modes which are coupled to two cavity modes $\hat d_1$ and   $\hat d_2$, and 
aim for the non-local dissipators given in Eq.(\ref{Eq.QuadDbasis1}). 
This implies that one mechanical mode realizes dissipative hopping between the cavity modes and the second mode a dissipative parametric amplification.
This can be realized by having two red detuned pumps at frequency $\omega_{1,m} - \omega_{1,2}$ for the hopping case
and  two blue detuned tones at  $\omega_{2,m} + \omega_{1,2}$ for the parametric amplification case. Note, it would be favorable that both cavity resonance frequencies
coincide, this could  reduce the number of pump sources to two.
Additionally, the two cavities are coupled via a coherent hopping interaction with strength $J$ and are coupled to input/output waveguides with strength $\kappa$.  
Working in an interaction picture with respect  to the free Hamiltonian we obtain (under a rotating wave approximation)
\begin{align}
 \hH = \lambda  \hat b_1^{\dag} \left( \hat d_{1} + \eta  \hat d_{2} \right)  + \lambda  \hat b_2^{\dag} \left( \hat d_{1}^{\dag}  +  \eta  \hat d_{2}^{\dag} \right) + J \hat d_1 \hat d_{2}^{\dag} + h.c.,
\end{align}
 where $\eta  $ are complex and account for an asymmetric coupling of the cavity modes to the mechanical modes. This is not crucial for the directionality in the system, but it will 
 allow us to optimize the noise properties of the resulting amplifier. 
As usual we assume that the mechanical modes are strongly damped and we can adiabatically eliminate them. This results in the equation of motion
for the cavity operators
\begin{align}
 \frac{d}{dt} \hat d_1 =&    - \sqrt{\kappa} \hat d_{1,\IN}  -   \frac{\kappa}{2} \hat d_1
                        + i   \frac{2\lambda  }{\sqrt{\gamma }} \hat b_{1,\IN}+ i   \frac{2\lambda  }{\sqrt{\gamma }} \hat b_{2,\IN}^{\dag}  
                        \nonumber \\ &
                        -   \left[   \frac{2\lambda ^2}{\gamma } \left(\eta   -   \eta^{\ast} \right)  + i J  \right] \hat d_2,
 \nonumber \\
  \frac{d}{dt} \hat d_2 =& - \sqrt{\kappa} \hat d_{2,\IN}  - \frac{\kappa}{2}  \hat d_2 
                           + i   \eta^{\ast}  \frac{2\lambda }{\sqrt{\gamma }} \hat b_{1,\IN} + i  \eta \frac{2 \lambda }{\sqrt{\gamma }} \hat b_{2,\IN}^{\dag} 
                           \nonumber \\ &
                           - \left[   \frac{2\lambda^2}{\gamma } \left( \eta^{\ast} -    \eta \right) + i J    \right]  \hat d_1  . 
\end{align}
In general, the coupling of two cavities to one engineered reservoir leads to local damping terms, which here cancel out as we couple to two reservoirs. 
This is an important property for the resulting bandwidth of the amplifier.
We aim for directional signal propagation from cavity 1 to cavity 2, thus we have the directionality condition
\begin{align}
 i  \frac{2\lambda^2}{\gamma }  \left(\eta   - \eta^{\ast} \right)   \equiv    J,  
\end{align}
which decouples cavity 1 from cavity 2. For optimal noise performance we  
set $\frac{4\lambda^2}{\gamma } \equiv \frac{\kappa}{4}$ and  define $\eta = \frac{i}{2} \sqrt{ \mathcal G_0} $,  which  leaves us with the output on resonance
 \begin{align}
 \hat d_{1,\OUT} =& -   \hat d_{1,\IN}  + i     \left(  \hat b_{1,\IN}+   \hat b_{2,\IN}^{\dag} \right)    ,
 \nonumber \\
 \hat d_{2,\OUT} =& -  \hat d_{2,\IN}  -  \sqrt{ \mathcal G_0}  \hat b_{2,\IN}^{\dag}    -   i  \sqrt{ \mathcal G_0}     \hat d_{1,\IN}  , 
\end{align}
this can be further optimized via impedance matching, which would require more free parameters, e.g., $\lambda_{1,2}$ instead of $\lambda$.
The frequency dependent gain and the added noise become
\begin{align}
 \mathcal G [\omega] =  \frac{  \mathcal G_0   }{ \left(1 + \frac{4\omega^2}{\kappa^2}   \right)^2 },
 \hspace{0.5cm}
 \bar n_{\textup{add}} = \frac{1}{2} + \bar n_{m,2}^T + \frac{1}{\mathcal G_0} \left(\frac{1}{2} + \bar n_{c,2}^T \right) .
 \label{eq:AddedNoiseDBasis}
\end{align}
Here we have the same frequency dependent gain as obtained in the quadrature basis; with the feature that
the bandwidth over which signals can be amplified is independent of the gain. On the other hand, the added noise differs slightly from the quadrature basis realization, cf.   Eq.~(\ref{Eq.:AddedNoiseQuadratureBasis}). Here only the noise contribution from one mechanical mode (i.e., one reservoir) shows up. However, both cases coincide if the reservoirs are at the same temperature, i.e., $\bar n_{r,1}^T = \bar n_{r,2}^T \equiv \bar n_{m,2}^T$ in Eq.~(\ref{Eq.:AddedNoiseQuadratureBasis}).


\end{document}